\documentclass[11pt]{article}
\usepackage[a4paper, portrait, margin=2.5cm]{geometry}
\usepackage{graphicx}
\usepackage{lastpage}
\usepackage{fancyhdr}

\fancypagestyle{firstpage}{%
  \setlength\headheight{154pt}
  \fancyhf{}%
  \fancyhead[C]{\includegraphics[width=\textwidth]{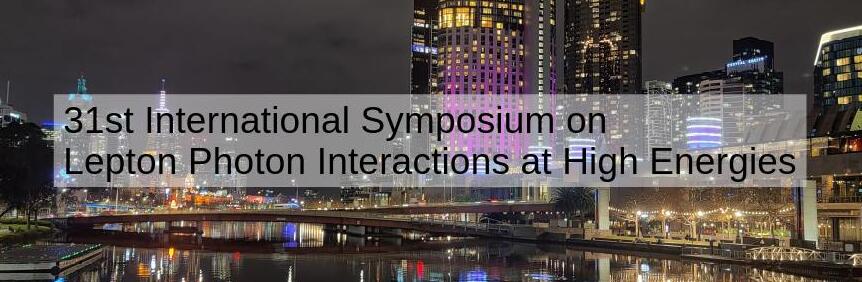}}%
  \fancyfoot[C]{CC-BY-4.0 licence}%
}

\begin{document}
\title{Future Challenges For Event Generators}
\date{17-21 July 2023}
\author{Davide~Napoletano \\ Universit\`{a} degli Studi di Milano-Bicocca}

\newgeometry{top=2cm, bottom=7cm}
\maketitle
\thispagestyle{firstpage}
\abstract{In this talk I present a personal perspective on what the current and
  future challenges are for Monte Carlo event generators. I focus in particular
  on those aspects of Monte Carlo event generators that have not, historically,
  received the same scrutiny and level of advancements which will be mandatory
  in future, cleaner and more precise experimental set-ups than current day
  LHC.}
\restoregeometry

\section{Introduction}
The main goal of a Monte Carlo event generator
(MCEG)~\cite{Sherpa:2019gpd,Sjostrand:2014zea,Bellm:2015jjp} is to describe, as
accurately as possible, physical events occurring in various experimental
set-ups\footnote{Please note that here and in the following MCEGs refers to
  general purpose generators. This means the references to providers of the
  underlying hard scattering process and matching are not listed here.}.
Even in the simplest of cases, this may look like an almost
impossible task, as an event consists of many interleaved effects that are
difficult to tackle all at once. Nevertheless, we have had great success in
comparing theoretical predictions obtained through MCEGs to experimental
data. The main drive for this success is the fact that while it is true that a
series of effects all take place in a physical event (such as ISR, MPI,
hadronisation, hadron decay) other than the hard scattering, these various
effects happen at different energies regimes (times), and can thus be considered
as independent from one another. The price one pays to make this approximation
can be estimated by scaling arguments, and one expects that the
correlation effects amongst these various aspects is suppressed by some power of
their characteristic energy scales with respect to that of the hard scattering.

The main strength of this approach is that it allows for a complete separation
of ingredients, where each of the separate building blocks can be improved
independently from one another, as long as they are then properly
matched. It is thus only natural that the main community effort has been devoted
to improving the accuracy of the hard scattering, which represents, from the
Monte Carlo perspective, the initial conditions that needs to be dressed with
all other ingredients. Among these ingredients there is the parton shower, which
incidentally is an aspect of Monte Carlo event generation that has received a
relatively large attention over the years. In the following, I quickly review
the main focus of theoretical developments relative to Monte Carlo event
generation, which in a sense represents the succeeded challenges the community
has faced. Then I describe what in my opinion are aspects that need to be
developed to a similar standard in order to succeed for the challenges
ahead. Lastly I conclude with some remarks on challenges the Monte Carlo
community specifically (and to some extent the broader high energy physics
community) needs to address in order not to repeat some of the mistakes that, in
my opinion, we have made. The main take home of this talk is that while it is
true that we can rely on factorisation, none of the single pieces of a Monte
Carlo work on their own when the aim is precision physics.

\section{Hard Scattering}
The description of the hard scattering process represents the core of
MCEGs. This is often referred to as fixed order, and, in general, momentum and
color information coming from the hard matrix element is then used to feed the
parton shower and hadronisation. As they represent the starting point of the
perturbative expansion -- in the sense that any other aspect of MCEG can be seen
as attaching higher orders in either the coupling constants or in powers of
$\Lambda_{\mathrm{QCD}}/Q$ to it -- it is only natural
that this is the aspect that has received the most attention over the years, and
has seen a lot of success. This success can essentially be split up in two
ingredients: calculation of higher order matrix elements and development of
subtractions. The former is relatively straightforward in its idea, while
technically complicated, and requires the calculation of multi-loop
diagrams. Indeed for one-loop calculations we now know that there exists a
finite set of ``base'' scalar integrals, and the only complication of performing
a one-loop calculation lies in finding the ``coefficient'' multiplying such base
integrals. Extensions of this to two- and higher-loops are not available yet in
full generality,
nevertheless we have now two-loops calculations for $2\to2,3$ scattering and
three-loops for $2\to1$ processes.

\begin{figure}[t]
  \begin{center}
    \includegraphics[width=0.8\textwidth]{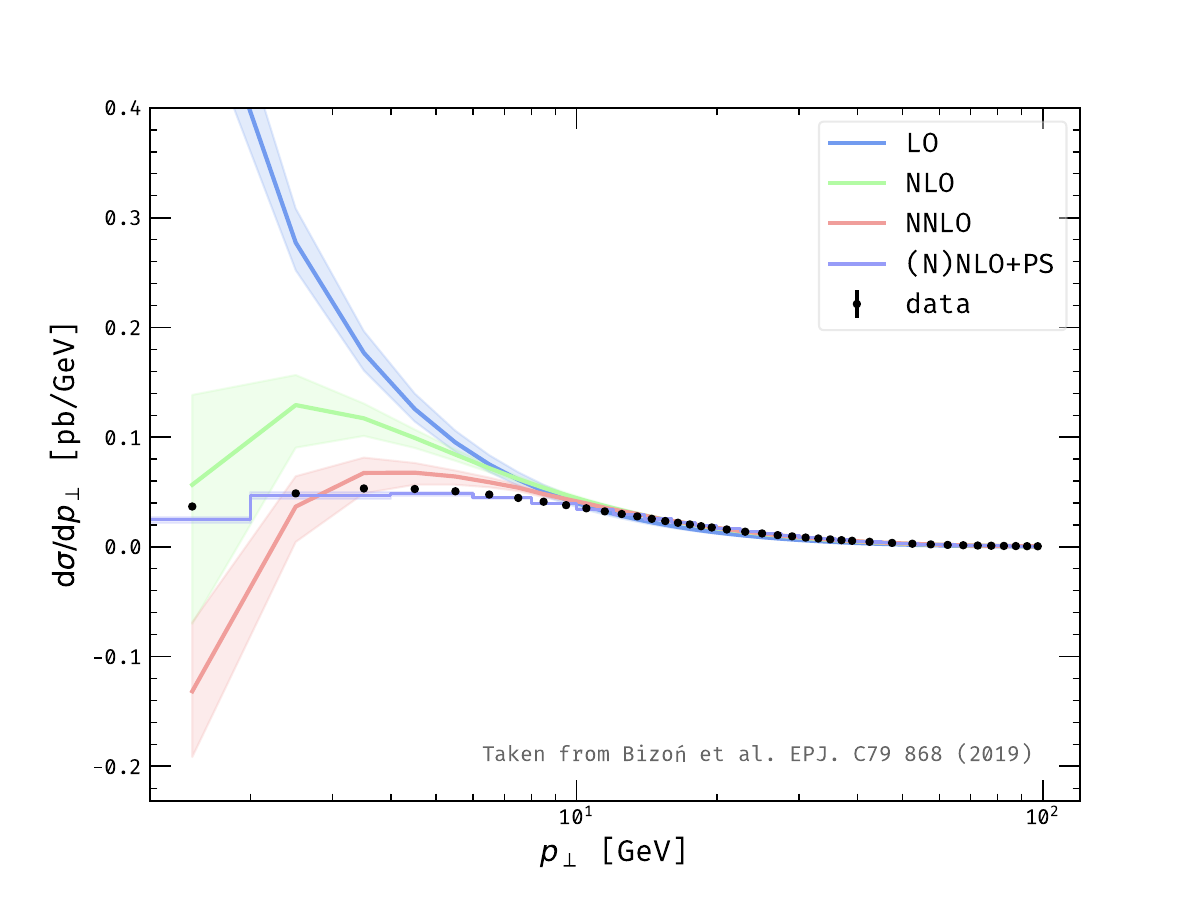}
    \caption{\label{fig:1} The importance of including some sort of evolution to
    low scales, in this case achieved by the parton shower, even when higher
    orders are included~\cite{Bizon:2019zgf}.}
  \end{center}
\end{figure}

Subtraction is, on the other hand, only a technical complication due to the
fact that the cancellation of infrared singularities (for both QCD and QED) does
not happen trivially, but only after the integration over the emission
phase-space, and computer programs cannot deal with this in an effective manner.
There is the additional complication that ideally one wants subtraction terms
that are both effective in subtracting the infrared singularities of the matrix
elements, and, at the same time, easy to integrate analytically, such as to
avoid numerical integration of a logarithmically enhanced (divergent)
phase-space.
These complications have, however, been overcome in a variety of approaches at
both NLO and NNLO. Equipped with both higher order calculations and subtraction,
we can try and compare to some experimental data. Take, $Z+j$ production in
Drell-Yan, for example, as depicted in Fig.~\ref{fig:1}.

As it can be seen, although the fixed order description captures well the
behaviour at large transverse momentum of the $Z$-boson, at low $p_\perp$ fixed
order alone in not enough, and one would need an infinite amount of higher orders
to accurately describe data. This is precisely what the parton shower does in
this case, and indeed one can see that after matching we recover a good
description of data across the entire spectrum.

\section{Parton Shower, Matching, Merging}
Parton showering can be seen in a variety of different ways. Here I present it in
the most pragmatic approach possible: it evolves -- by {\it radiating}
-- particles produced in the hard scattering to lower energies, energies closer
to the hadronisation scale. Note that I am being purposefully vague here as
describing this process in detail requires much more time and space than a
proceeding allows. The core idea behind parton showers dates back to the
'80s~\cite{Fox:1979ag},
and has to a large extent remain unchanged since. It consists of dressing with
radiation, ordered in a suitably defined variable such as to reflect time
ordering, the hard matrix element. The implementation details of this vary to a
large extent, as one can envision angular ordering, energy ordering, invariant
mass ordering, transverse momentum (in various definition) ordering and so on. On
top of this, how exactly the kinematics of the splittings is implemented, at what
scale the coupling attached to a given splitting is evaluated all constitute the
``implementation details'' of a given shower algorithm.

For about 20 or so years, it was believed that no matter what one did with most
of these choices, as long as you had an algorithm capable of describing
coherence, you evaluated $\alpha_s$ in the so-called CMW scheme, with leading
order splitting functions, you would get at least a leading-logarithmic accurate
description of all observables, which could even be next-to-leading logarithmic
accurate for a special class of them. This belief was dismantled when it was
shown that, due exactly to those unimportant implementation details, dipole
showers break NLL accuracy and LL accuracy beyond leading colour. This has
started a new series of more accurate shower algorithms, and the aim is to be
able achieve NNLL for most observables.

\begin{figure}[t]
  \begin{center}
    \includegraphics[width=0.8\textwidth]{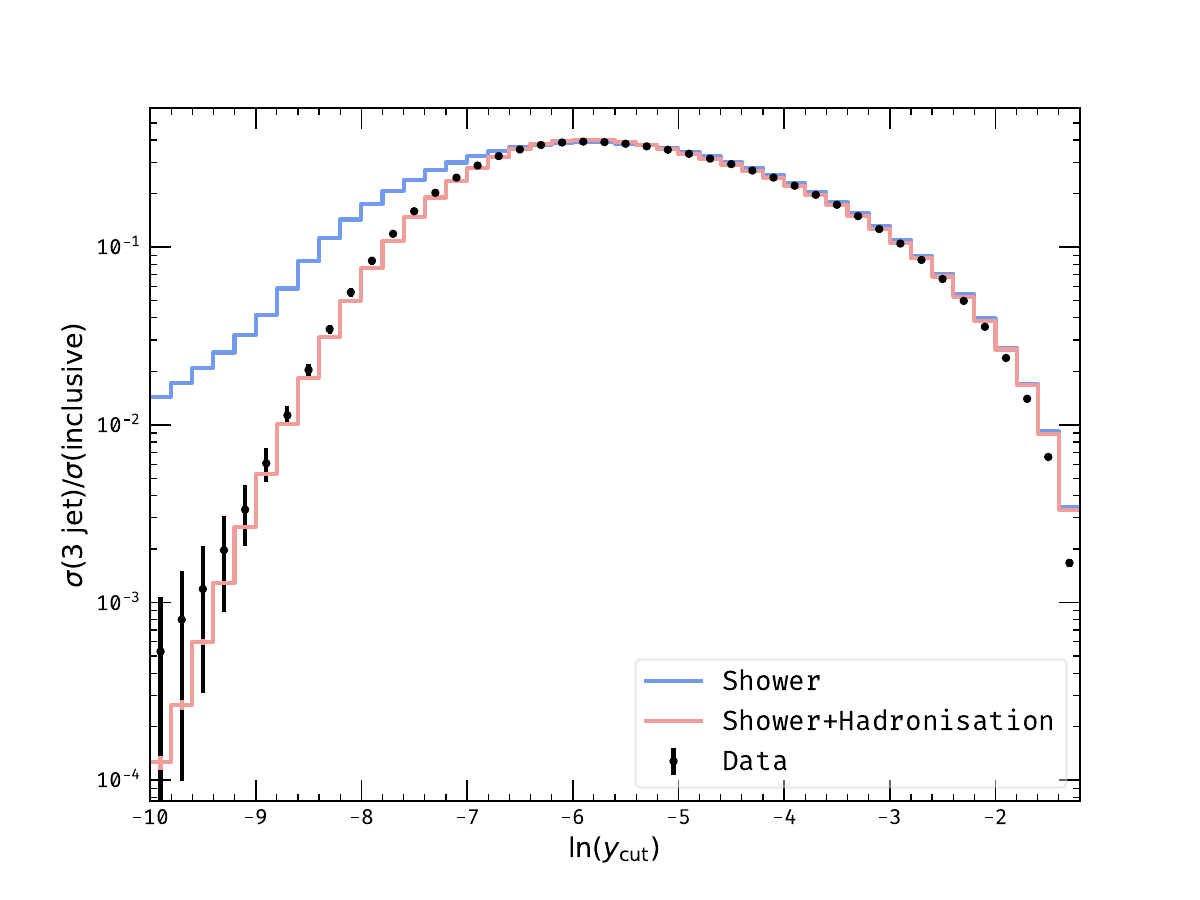}
    \caption{\label{fig:2} The importance of including non-perturbative
      corrections, even when shower and fixed-order corrections are included.}
  \end{center}
\end{figure}

Nevertheless, based on the discussion around Fig~\ref{fig:1} one would expect
that the shower description would, by itself, give a fairly good description of
data. However, when comparing for example to LEP data, as done in
Fig.~\ref{fig:2}, one can see that to actually match data across the entire
spectrum, one needs to include hadronisation and subsequent fragmentation
effects, on top of shower emissions. Note that this goes beyond the accuracy of
the shower, as the physics behind these two regimes is intrinsically different
-- one is perturbative and the other non-perturbative -- thus the same argument
can be equally applied to more accurate showers. However there is a non-trivial
subtlety in Fig.~\ref{fig:2} and the following discussion. Hadronisation models,
and more in general most models involving the non-perturbative side of event
generation, undergo tuning, {\it i.e.} they match data because they were fitted
so.

\section{Non-Perturbative models}
The core idea behind hadronisation models~\cite{Andersson:1983jt,Webber:1983if}
and more broadly non-perturbative models used in MCEGs, is that they are
phenomenological models based on a --
tipically large -- number of parameters that control various aspects of
non-perturbative physics, such as the mean and distribution of charged particle
multiplicities, the yield of individual hadron species, the fragmentation of
heavy quarks to heavy hadrons and so on. Different models, such as ``cluster''
or ``strings'', eventually differ in how partons, coming from the parton shower,
are combined together to form hadrons and their subsequent decays products, but
they share the main logic explained above.

The hope, and the common belief -- although no proof or disproof exists as of
yet -- is that, similarly to parton distribution functions (PDF) which are extracted
from DIS data and used for any process requiring them, one can tune these
parameters in a clean environment such as LEP, and then, provided they are
universal to some extent, re-use them for other set ups. However, one can see
how this might fail. Indeed take the PDF example: they are essentially one
dimensional functions of a given parton's momentum fraction, $x$, and that is
the only non-perturbative part of these functions. Their scale dependence is
completely determined perturbatively -- a.k.a. DGLAP equations. This means that
the dependence on the point (factorisation scale) where one decides arbitrarily
to divide the non-perturbative and perturbative regions is completely known and
cancels between hard partonic matrix elements and PDF evolution at any fixed-order.
In contrast, non-perturbative models' parameters have no scale dependence --
apart from that induced by tuning them at a given scale -- and their evolution
is not known theoretically, as such it is easy to imagine that extracting them at a
given scale would not give the same result as extracting them at another energy
scale, or in another experimental set-up. Indeed, novel tunings based on H1 data
and clean DIS observables are starting to show evidence of incompatibility with
LEP data tunes.

An additional issue is that although higher order calculations are available
tunes that include
such perturbative effects as to try and reduce as much as possible tuning away
higher order effects are still scarce.
All in all, given that -- as seen for example in Fig.~\ref{fig:2} --
non-perturbative effects can give extremely large corrections in some regions of
phase-space, we need to be able to at least understand these models better to
better asses uncertainties on theoretical calculations.

\section{Miscellaneous and Conclusions}
As MCEGs cover all aspects of a collider event, and that the majority of people
doing research on these topics focuses on either the hard matrix elements or the
shower and their interplay, it is easy to imagine that the
list of understudied topics can become quite long. Here I list a few of them,
keeping in mind that this does not have the aim of being an exhaustive and
complete list by any mean.
\paragraph{Heavy Quark Mass effects}
One important aspect of Monte Carlo simulations that needs to be theoretically
under control is the inclusion of heavy quark mass effects, specifically in parton
showers. Historically the approach has been that of replacing massless with
massive ingredients and hope for the best, with the idea that mass effects are
in any case beyond the current accuracy claimed by parton showers. However,
there are two important aspects to note here. First, while this may be true in
general, as we move towards more exclusive observables there is a non-trivial
scale interplay between the mass of the heavy quark and the energy -- in the
broader sense -- of a given observable, which may lead to large effects which
have to be controlled. Second, as the accuracy of parton showers increases,
these have to be included to claim the full higher order accuracy claimed. In
addition to these two aspects, there is the subtle thread of how to estimate an
uncertainty of a Monte Carlo simulation. If we trust that including or not mass
effects is only a matter of factorisation scheme dependence -- up to purely
kinematical effects of course -- then the difference between including them or not,
or how these are included, should be treated as a pure theoretical uncertainty,
which would lead to much larger uncertainties that the ones currently reported.
\paragraph{Electroweak Corrections}
Electroweak corrections have had, in recent years, an important
role. Implementations of fully automated subtractions at NLO have been
implemented in most MCGs and one-loop matrix elements from loop providers are
readily available. In addition, a variety of implementations of these
corrections in approximations were also implemented and are now fully automated,
including their matching to QCD higher orders and parton shower, including
merging.
There are however two important aspects, in my opinion, that have not seen the
same level of attention, and that I think need to be considered amongst the set
of future challenges. The first is the availability -- theory papers on the
topic and implementation exist but it is not entirely clear in which shape these
are -- of EW parton showers. The second, which is tied to the first in some
sense, is the need to revisit our current concept of what an EW final state
really is. This is a potentially long point to elaborate in full detail, but in
short, when talking about EW corrections we typically only refer to virtual
corrections, as the argument is that we can, experimentally, distinguish between
a $Z$ boson or a photon or a $W$ boson and a final state with or without an
additional massive vector boson. This is certainly true at the LHC, but to what
extent this remains true at higher energies -- where virtual electroweak
corrections would play a relatively bigger role -- remains to be seen. The point
is that the inclusion of real radiation at high energies scales almost exactly
like the virtuals with opposite sign, leaving only a mass-suppressed
mis-cancellation. At the same time, even if it remains true that only virtual EW
corrections contribute, at higher energies we need to developed frameworks
capable of handling the resummation of Sudakov logarithms.

\paragraph{Conclusions}
Monte Carlo event generators are a fundamental tool to compare theoretical
predictions to experimental data, as a theoretical tool for phenomenological
studies. Most of the successful challenges of the past have been dedicated to
develop technologies to be able to include higher order calculations and to
match those with parton showers. More recently, parton showers have received a
lot of attention which is thus leading towards having more accurate parton
showers. Technologies on how to then match higher order accuracies parton
showers with higher order calculations will have to be then developed, and in
some sense this will likely be a crucial not-so-distant-future challenge.
On top of this, a variety of non-perturbative and/or power-suppressed aspects of
Monte Carlo event generators need to be put on more solid theoretical
grounds. This is important both to provide more accurate results, and because
by developing these aspects allows us to estimate theoretical
uncertainties in a fairer way.

In order for the future of Monte Carlo Event Generators to remain as bright as
it is been, on top of physics
challenges, I personally think we have to tackle more pragmatic ones as
well. In general codes developed for wider uses, such as MCEGs, are huge and
highly complicated pieces of code which
not only take a long time to write and maintain, but also to learn and master to
a point where developments are possible. In addition, their broad use requires
that developers invest a lot of their work time in maintaining the code and
supporting users. Our current way of
evaluating scientific success, based on pure metric, does not help in this sense.
New ideas take a lot of time to be thought of, and even more
time to be transformed into practical algorithms, and as such are
disfavored. Furthermore neither maintenance nor user support lead to
publications, which are the only way to get positions and grants.
To successfully tackle this future
challenges,  which will require even more new ideas on the one hand, and more
advanced coding and maintenance and support on the other hand, we need to re-assess
as a community how to measure in a fairer way the scientific value of people's
work.

\bibliographystyle{unsrt}
\bibliography{MonteCarlo}

\begin{thebibliography}{1}

\bibitem{Sherpa:2019gpd}
Enrico Bothmann et~al.
\newblock {Event Generation with Sherpa 2.2}.
\newblock {\em SciPost Phys.}, 7(3):034, 2019.

\bibitem{Sjostrand:2014zea}
Torbj\"orn Sj\"ostrand, Stefan Ask, Jesper~R. Christiansen, Richard Corke,
  Nishita Desai, Philip Ilten, Stephen Mrenna, Stefan Prestel, Christine~O.
  Rasmussen, and Peter~Z. Skands.
\newblock {An introduction to PYTHIA 8.2}.
\newblock {\em Comput. Phys. Commun.}, 191:159--177, 2015.

\bibitem{Bellm:2015jjp}
Johannes Bellm et~al.
\newblock {Herwig 7.0/Herwig++ 3.0 release note}.
\newblock {\em Eur. Phys. J. C}, 76(4):196, 2016.

\bibitem{Bizon:2019zgf}
Wojciech Bizon, Aude Gehrmann-De~Ridder, Thomas Gehrmann, Nigel Glover,
  Alexander Huss, Pier~Francesco Monni, Emanuele Re, Luca Rottoli, and
  Duncan~M. Walker.
\newblock {The transverse momentum spectrum of weak gauge bosons at N ${}^3$ LL
  + NNLO}.
\newblock {\em Eur. Phys. J. C}, 79(10):868, 2019.

\bibitem{Fox:1979ag}
Geoffrey~C. Fox and Stephen Wolfram.
\newblock {A Model for Parton Showers in QCD}.
\newblock {\em Nucl. Phys. B}, 168:285--295, 1980.

\bibitem{Andersson:1983jt}
Bo~Andersson, G.~Gustafson, and B.~Soderberg.
\newblock {A General Model for Jet Fragmentation}.
\newblock {\em Z. Phys. C}, 20:317, 1983.

\bibitem{Webber:1983if}
B.~R. Webber.
\newblock {A QCD Model for Jet Fragmentation Including Soft Gluon
  Interference}.
\newblock {\em Nucl. Phys. B}, 238:492--528, 1984.

\end{thebibliography}

\end{document}